\documentclass[prd,twocolumn,showpacs,amsmath,amssymb]{revtex4-1}
\usepackage{amssymb}
\usepackage{amsfonts}
\usepackage{overpic,graphicx}
\usepackage{keyval,graphicx}
\usepackage{textcomp,wasysym}
\usepackage[perpage,symbol]{footmisc}%
\usepackage{lineno}
\usepackage[dvipdfm,CJKbookmarks=true, colorlinks=true,
linkcolor=blue, urlcolor=blue,citecolor=blue]{hyperref}
\usepackage{ulem}

\begin{document}

\title{\boldmath Search for C-parity violation in $J/ \psi \to \gamma\gamma$ and $ \gamma \phi$}

\author{
\begin{small}
\begin{center}
M.~Ablikim$^{1}$, M.~N.~Achasov$^{8,a}$, X.~C.~Ai$^{1}$,
O.~Albayrak$^{4}$, M.~Albrecht$^{3}$, D.~J.~Ambrose$^{42}$,
A.~Amoroso$^{46A,46C}$, F.~F.~An$^{1}$, Q.~An$^{43}$, J.~Z.~Bai$^{1}$
% ...
\\
\vspace{0.2cm}
(BESIII Collaboration)\\
\vspace{0.2cm} {\it
$^{1}$ Institute of High Energy Physics, Beijing 100049, People's
Republic of China\\
% ...
$^{3}$ Bochum Ruhr-University, D-44780 Bochum, Germany\\
$^{4}$ Carnegie Mellon University, Pittsburgh, Pennsylvania 15213, USA\\
% ...
$^{8}$ G.I. Budker Institute of Nuclear Physics SB RAS (BINP),
Novosibirsk 630090, Russia\\
% ...
$^{42}$ University of Rochester, Rochester, New York 14627, USA\\
$^{43}$ University of Science and Technology of China, Hefei 230026,
People's Republic of China\\
% ...
$^{46}$ (A)University of Turin, I-10125, Turin, Italy; (B)University of
Eastern Piedmont, I-15121, Alessandria, Italy; (C)INFN, I-10125, Turin,
Italy\\
% ...
$^{a}$ Also at the Novosibirsk State University, Novosibirsk, 630090,
Russia\\
% ...
}\end{center}
\vspace{0.4cm}
\end{small}
}

\affiliation{}

\begin{abstract}
Using $1.06\times10^8$ $\psi(3686)$ events recorded in $e^{+}e^{-}$ collisions at $\sqrt{s}=$ 3.686 GeV
with the BESIII at the BEPCII collider, we present searches for C-parity
violation in $J/\psi \to \gamma\gamma$ and $ \gamma \phi$ decays via $\psi(3686) \to J/\psi \pi^+\pi^-$.
No significant signals are observed in either channel.
Upper limits on the branching fractions
are set to be $\mathcal{B}(J/\psi \to \gamma\gamma) < 2.7 \times
10^{-7}$ and $\mathcal{B}(J/\psi \to \gamma\phi)  < 1.4 \times
10^{-6}$ at the 90\% confidence level. The former is one order of
magnitude more stringent than the previous
upper limit, and the latter
represents the first limit on this decay channel.
\end{abstract}

\maketitle

\oddsidemargin -0.2cm
\evensidemargin -0.2cm

\section{\bf INTRODUCTION}

The charge conjugation~(C) operation transforms a particle into its
antiparticle and vice versa.  In the Standard Model~(SM), C invariance
is held in strong and electromagnetic~(EM) interactions.
Until now, no C-violating processes have been observed in EM
interactions~\cite{pdg14}. While both C-parity and P-parity can be
violated in the weak sector of the electroweak interactions in the SM,
evidence for C violation in the EM sector would immediately indicate
physics beyond the SM.

Tests of C invariance in EM interactions have been carried out by many
experiments~\cite{pdg14}. In $J/\psi$ decays, however, only the
channel $J/\psi \to \gamma\gamma$ has been
studied~\cite{jpsic1,jpsic2,jpsic3,jpsic4}, and the corresponding best
upper limit on the branching fraction is $5\times 10^{-6}$, measured
by the CLEO Collaboration.
In this paper, we report on searches for the decays of $J/\psi \to
\gamma\gamma $ and $\gamma\phi$ via $\psi(3686) \to J/\psi
\pi^+\pi^-$.
The analysis is based on a data sample corresponding to
$1.06\times10^8$ $\psi(3686)$ events collected at $\sqrt{s}=$ 3.686
GeV (referred to as on-resonance data)~\cite{npsip} and a data set of
44.5~pb$^{-1}$ collected at 3.650~GeV (referred to as off-resonance
data)~\cite{continuumdata} with the Beijing Spectrometer~(BESIII).

\section{BESIII AND BEPCII}

The BESIII detector at the BEPCII~\cite{bes3} double-ring $e^+e^-$ collider
is a major upgrade of the BESII
experiment at the Beijing Electron-Positron Collider~(BEPC)~\cite{bepc2}
for studies of physics in the $\tau$-charm energy region~\cite{bes3physics}. The design peak luminosity of
BEPCII is $10^{33}$~cm$^{-2}$~s$^{-1}$ at
a beam current of 0.93~A.  Until now, the achieved peak luminosity is $7.08\times10^{32}$~cm$^{-2}$~s$^{-1}$
at 3773~MeV. The BESIII detector, with a
geometrical acceptance of 93\% of 4$\pi$, consists of the
following main components. (1)~A small-celled
main drift chamber~(MDC) with 43 layers is used to track charged particles. The average
single-wire resolution is 135~$\mu$m, and the momentum
resolution for 1~GeV/$c$ charged particles in a 1~T magnetic
field is 0.5\%. (2)~An EM calorimeter~(EMC) is used to measure photon energies.
The EMC is made of 6240 CsI~(Tl) crystals arranged in a cylindrical
shape (barrel) plus two end caps. For 1.0~GeV photons, the
energy resolution is 2.5\% in the barrel and 5\% in the end-caps,
and the position resolution is 6~mm in the barrel and
9~mm in the end caps. (3)~A time-of-flight system~(TOF) is used for
particle identification.  It is composed of a barrel made of
two layers, each consisting of 88 pieces of 5~cm thick and 2.4~m long plastic
scintillators, as well as two end-caps with 96 fan-shaped,
5~cm thick, plastic scintillators in each end cap.
The time resolution is 80~ps in the barrel and 110~ps in the
end caps, providing a $K/\pi$ separation of more than
2$\sigma$ for momenta up to about 1.0~GeV/$c$. (4)~The muon chamber
system is made of resistive plate chambers
arranged in 9 layers in the barrel and 8 layers in the end-caps
and is incorporated into the return iron yoke of the superconducting
magnet. The position resolution is about 2~cm.

The optimization of the event selection and the estimation
of background contributions from $\psi(3686)$ decays are performed through
Monte Carlo~(MC) simulations. The {\sc GEANT4}-based simulation
software {\sc BOOST}~\cite{boost} includes the geometric and
material description of the BESIII detectors, the detector
response and digitization models, as well as a record of
the detector running conditions and performances. The
production of the $\psi(3686)$ resonance is simulated by
the MC event generator {\sc KKMC}~\cite{kkmc}, while the
decays are generated by {\sc EVTGEN}~\cite{evtgen} for known decay
modes with branching ratios being set to the PDG~\cite{pdg10}
world average values, and by {\sc LUNDCHARM}~\cite{lundcharm} for the
remaining unknown decays. The process of $\psi(3686) \to J/\psi \pi^+\pi^-$ is
generated according to the formulas and measured results in Ref.~\cite{bes2jpipi},
which takes the small D-wave contribution into account. The signal channels, $J/\psi
\to\gamma\gamma$ and $\gamma\phi$, are generated
according to phase space.
The process $\phi \to K^+K^-$ is generated using a $\sin^2\theta$
distribution, where $\theta$ is the helicity angle of the kaon defined
in the $\phi$ center-of-mass system. To obtain upper limits from
the measured distributions, we test both
the Bayesian method~\cite{bayesian} and the Feldman-Cousins construction~\cite{feldman} and choose for
each channel the method resulting in the most stringent upper limit.

\section{Search for \mbox{\boldmath$J/\psi \to \gamma\gamma$}}
\label{sec:gg}
To search for $J/\psi \to\gamma\gamma$ via $\psi(3686) \to J/\psi \pi^+\pi^-$, candidate events with the topology
$\gamma\gamma\pi^+\pi^-$ are selected using the following criteria.
For each candidate event, we require that at least two charged tracks are
reconstructed in the MDC and that the polar angles of the tracks satisfy
$|\cos\theta|<0.93$. The tracks are required to
pass within $\pm10$~cm of the interaction point along the
beam direction and within $\pm1$~cm in the plane perpendicular
to the beam. Photon candidates are reconstructed by clusters of energy deposited in the EMC.
The energy deposited in the TOF counter in front of the EMC is included to improve the
reconstruction efficiency and the energy resolution. Photon candidates are required to have
deposited energy larger than 25 MeV in the barrel region ($|\cos\theta| < 0.80$) or 50 MeV
in the end-cap region ($0.86 <|\cos\theta| < 0.92$). Showers on the edge of the barrel and end-caps are poorly measured
and are excluded. EMC cluster timing requirements ($0\leq t \leq 14$ in units of 50~ns) are
used to suppress electronic noise and energy deposits unrelated
to the event. Only events with exactly two photon candidates are
retained for further analysis. In addition, the energies of both photons are required to be greater than 1.0~GeV.

Two oppositely charged tracks, with momentum less than 0.45 GeV/$c$,
are selected and assumed to be pions without particle identification.
We impose $|\cos\theta_{\pi^+\pi^-}|<0.95$ to exclude random
combinations and reject backgrounds from $e^+e^- \to \gamma\gamma
e^+e^-$ events, where $\theta_{\pi^+\pi^-}$ is the angle between the
two oppositely charged tracks.

A kinematic fit enforcing energy-momentum conservation is performed under the $\gamma\gamma\pi^+\pi^-$
hypothesis, and the obtained $\chi^2_{\rm 4C}$ value of the fit is required to be $\chi^2_{\rm 4C}<40$
to accept an event for further analysis. After applying the previous selection
criteria, only one combination is found in
each event, both in data and simulation.

\begin{figure}[htbp]
\begin{center}
\includegraphics[width=8cm]{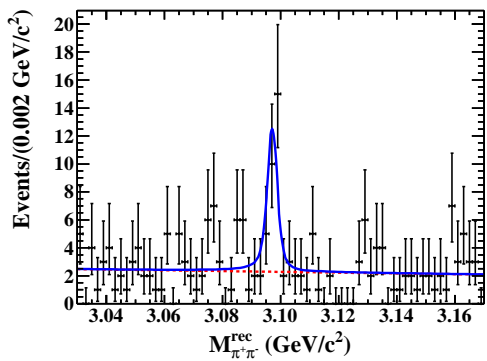}
\caption{\small The $ M^{\rm rec}_{\pi^+\pi^-}$ (calculated from MDC measurements) distribution for
$\psi(3686) \to J/\psi \pi^+\pi^-, J/\psi \to \gamma\gamma$
candidate events from on-resonance data. The solid curve shows the
global fit results and the dashed line indicates the non-peaking backgrounds.}\label{fig:data:gg}
\end{center}
\end{figure}

The candidate signal events are studied by examining the invariant mass recoiling
against $\pi^+\pi^-$, $M^{\rm rec}_{\pi^+\pi^-}$,
which is calculated using the momentum vectors of the corresponding tracks measured in the MDC. Figure \ref{fig:data:gg} shows the
resulting distribution of $M^{\rm rec}_{\pi^+\pi^-}$ from the
candidates for $\psi(3686) \to J/\psi \pi^+\pi^-, J/\psi \to \gamma\gamma $ from on-resonance data.
A $J/\psi$ signal is clearly observed, which,
as indicated by the studies described later, is dominated by backgrounds.
The $M^{\rm rec}_{\pi^+\pi^-}$ spectrum is fitted using an unbinned maximum likelihood fit. The $J/\psi$ signal line shape
is extracted from a control sample, $\psi(3686) \to J/\psi \pi^+\pi^-, J/\psi \to \mu^+\mu^-$,
selected from the on-resonance data. A first-order Chebychev polynomial is used to describe the non-peaking background. The fit determines the number
of observed events to be $N^{\rm obs} = 29.2\pm7.1$.

The main peaking backgrounds come from $\psi(3686) \to J/\psi
\pi^+\pi^-, J/\psi \to \gamma \pi^0, \gamma\eta, \gamma\eta_c$ and
$3\gamma$ ($\pi^0/\eta /\eta_c \to \gamma\gamma$). Large exclusive MC
samples are generated to study the peaking backgrounds, where $J/\psi
\to \gamma \pi^0$ and $\gamma\eta$ are generated by
the HELAMP generator of EVTGEN~\cite{evtgen}
to model the angular distribution; the other exclusive MC samples are generated
according to phase space. The same signal extraction procedure
is performed on each exclusive MC sample. Then the
contribution of each individual process is estimated by normalizing
the yields separately according to the equivalent generated
luminosities and the branching fractions taken from the PDG~\cite{pdg14}.
The normalized number of background events for the peaking
backgrounds are summarized in Table \ref{tab:ggbkg}.
Contributions
from other background channels such as $J/\psi \to \gamma f_2, f_2 \to
\pi^0\pi^0$ and $J/\psi \to \gamma \eta', \eta' \to \pi^0\pi^0\eta,
\eta \to \gamma\gamma$ are negligible. The backgrounds from continuum
processes are studied with the off-resonance data.  No peaking
background is identified from those. Summing up the contributions of the individual channels,
we obtain a total of $45.3 \pm 2.5$ expected peaking background events (see
Table \ref{tab:ggbkg}).

Since the two decay channels $J/\psi \to \gamma \pi^0$ and
$J/\psi \to \gamma \eta$ are expected to yield the dominant
contribution to the peaking background, we perform further studies on
these channels.  We examine the branching fractions with 106 M
simulated inclusive $\psi(3686)$ events and find good agreement
between the branching fractions used as input to the simulation and
the one measured on this MC sample.  We also roughly measure the
branching fractions of both channels with the same data set and find
results consistent with those listed at PDG~\cite{pdg14}.
The smooth backgrounds visible in Fig.~\ref{fig:data:gg} are also
reasonably well described by the background sources mentioned above.
These studies indicate that the above background estimation is
reliable.

\begin{table}[htbp]
\begin{center}
\caption{\small The expected number of peaking background events
  ($N^{\rm bkg}$) for $J/\psi \to \gamma\gamma$. The uncertainties
include the statistical uncertainty and uncertainty of all
intermediate resonance decay branching fractions.}\label{tab:ggbkg}
\begin{small}
\begin{tabular}{lcc}\hline
Background channel &     &  Expected counts ($N^{\rm bkg}$)  \\ \hline
$J/\psi \to \gamma\pi^0,\pi^0\to 2\gamma $  &      &$18.5\pm1.9$ \\
$J/\psi \to \gamma\eta ,\eta\to 2\gamma $  &   &$24.6\pm1.6$ \\
$J/\psi \to \gamma\eta_c,\eta_c\to 2\gamma $ &       & $1.3\pm0.3$ \\
$J/\psi \to 3\gamma $      &  &$0.9\pm0.3$  \\ \hline
Total                      &   & $45.3 \pm 2.5$ \\
\hline
\end{tabular}
\end{small}
\end{center}
\end{table}

After subtracting the background events from the total yields, we
obtain the net number of events as $N^{\rm net} =-16.1\pm7.5$.
Both methods to
obtain upper limits are tested, and the Feldman-Cousins method,
the one resulting in a more stringent upper limit, is chosen.
According to the Feldman-Cousins method, assuming a
Gaussian distribution and constraining the net number to be
non-negative, the upper limit on the number of $J/\psi \to
\gamma\gamma$ events is estimated to be $N^{\rm up}_{\rm sig} = 2.8$
at the 90\% confidence level (C.L.).

\section{Search for \mbox{\boldmath$J/\psi \to \gamma\phi$}}

To search for $J/\psi \to \gamma\phi$ via $\psi(3686) \to J/\psi \pi^+\pi^-$, candidate events with the topology
$\gamma K^+ K^- \pi^+\pi^-$ are selected using the following criteria.
The selection criteria for charged tracks and photons are the same as those listed in Section~\ref{sec:gg}.
Candidate events must have four charged tracks with
zero net charge and at least one photon with energy greater than 1.0~GeV.
The selection criteria for $\pi^+\pi^-$ are the same as
before except that we require $\cos\theta_{\pi^+\pi^-}<0.95$ in this case to exclude random combinations.

For other charged particles, the particle identification (PID)
confidence levels are calculated from the $dE/dx$ and time-of-flight
measurements under a pion, kaon or proton hypothesis. For kaon
candidates, we require that the confidence level for the kaon
hypothesis is larger than the corresponding confidence levels for the
pion and proton hypotheses. Two kaons with opposite charge are
required in each candidate event.

All combinations of the four charged tracks with
one high energetic photon are subjected
to a kinematic fit imposing energy-momentum conservation.
Candidates with $\chi^2_{\rm 4C}<40$ are accepted. If more than one
combination from photons satisfies the selection criteria in an event,
only the combination with the minimum $\chi^2_{\rm 4C}$ is retained.
Finally, only events are retained in which the mass recoiling against
the di-pion system satisfies
$3.082<M^{\rm rec}_{\pi^+\pi^-}<3.112$ GeV/$c^2$.

\begin{figure}[htbp]
\begin{center}
\includegraphics[width=8cm]{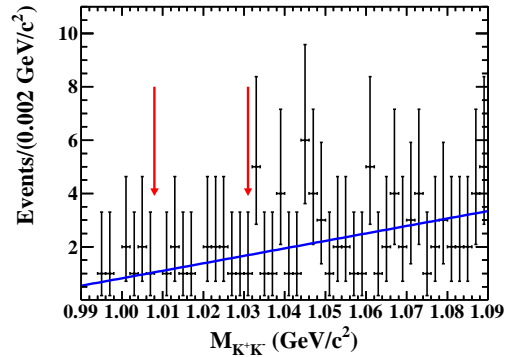}
\caption{\small The $ M_{K^+K^-}$ distribution for $\psi(3686) \to J/\psi \pi^+\pi^-, J/\psi \to \gamma\phi, \phi \to K^+K^-$
candidate events from on-resonance data. The solid line shows the
global fit results and the dashed line shows the background, and they are overlap each other.
The region between the arrows contains about 90\% of the signal according
to MC simulation.}\label{fig:data:gphi}
\end{center}
\end{figure}

The candidate signal events are studied by examining the invariant
$K^+K^-$ mass, $M_{K^+K^-}$, where the momenta obtained from the
kinematic fit are used to improve the mass resolution.
Figure~\ref{fig:data:gphi} shows the resulting $M_{K^+K^-}$ spectrum for
$\psi(3686) \to J/\psi \pi^+\pi^-, J/\psi \to \gamma\phi,\phi \to
K^+K^- $ candidates selected from on-resonance data.

An unbinned maximum likelihood fit is performed to extract the number
of reconstructed candidate events from the $K^+K^-$ invariant-mass
spectrum. The $\phi$ signal line shape is extracted from a MC
simulation. A first order Chebychev polynomial is used to describe the
background, which is shown in Fig.~\ref{fig:data:gphi}. The fit
yields $0.0\pm4.6$ events.

An MC study shows that there are no peaking background contributions.
The main possible non-peaking backgrounds come from $\psi(3686) \to
J/\psi \pi^+\pi^-, J/\psi \to \gamma f_2(1270), \pi^0K^+K^-$ and $
\pi^0 a^0_2$.  There are no candidates from the off-resonance
data observed; we therefore neglect the contribution from continuum
processes.

To obtain the upper limit, both methods are tested and in
this case the Bayesian method is chosen.
We determine the upper limit on the observed number of
events ($N_{\rm sig}^{\rm up}$) with the Bayesian method at the 90\% C.L. as
\begin{equation}
\frac{\int_0^{N_{\rm sig}^{\rm up} }\mathcal{L} dN_{\rm sig} }{\int_0^{\infty }
\mathcal{L} dN_{\rm sig}} = 0.90, \nonumber
\end{equation}
where $\mathcal{L}$ is the value of likelihood as a function of $N_{\rm sig} $.
The upper limit on the number of $J/\psi \to \gamma\phi$ is determined to be 6.9.

\section{Systematic Uncertainties}
\label{syserr}

The systematic uncertainties in the
measurements are summarized in Table~\ref{tab:syserr}.

The uncertainties in the tracking efficiency and kaon identification have been
studied in Ref.~\cite{trkpid}, which are 2.0\% per track and 2.0\% per kaon, respectively.

The energies of the photons in both channels are greater than 1.0~GeV. The uncertainty due to
the detection efficiency of high energy photons is estimated to be less than 0.25\% using
$J/\psi \to \gamma\eta'$, described in Ref.~\cite{photon}. We therefore assign 0.25\% per
photon as the systematic uncertainty for photon detection.

The uncertainty of the kinematic fit for the $J/\psi \to \gamma\gamma$ channel is estimated
from a control sample of $\psi(3686) \to \gamma\eta',\eta' \to \gamma\rho^0, \rho^0 \to \pi^+\pi^-$.
The efficiency is obtained from the change in the yield of $\eta'$ signal by
a fit to the $\gamma\pi^+\pi^-$ invariant-mass spectrum with or without the
requirement of $\chi^2_{\rm 4C} <40$ of the kinematic fit. The systematic uncertainty is determined to
be 1.9\%. The uncertainty of the kinematic fit for the $J/\psi \to \gamma\phi$ channel
is estimated to be 3.5\% from $\psi(3686) \to \gamma\chi_{\rm cJ}, \chi_{\rm cJ}
\to K^+K^- \pi^+\pi^-$.

The uncertainty associated with the
requirement on the number of good photons ($N_\gamma$)
for the $J/\psi \to \gamma\gamma$ channel is estimated by
using a control sample of $\psi(3686) \to J/\psi \pi^+\pi^-, J/\psi
\to \gamma\eta, \eta \to \gamma\gamma$ events. The differences of
selection efficiencies with and without the $N_\gamma$ requirement
($N_\gamma=3$ for the control sample) between data and MC is 3.0\%, which
is taken as the systematic uncertainty due to the $N_\gamma$ requirement.

By comparing the differences of selection efficiencies with and
without the $\cos\theta_{\pi^+\pi^-}$ requirement between data and MC,
the uncertainties due to this requirement for both channels are
estimated to be 0.9\% and 0.8\%, respectively.

The uncertainty due to the requirement of $M^{\rm rec}_{\pi^+\pi^-}$
to be within the $J/\psi$ signal region for $J/\psi \to \gamma \phi$
is estimated as 1.4\% by comparing the selection efficiencies between
data and MC.

The uncertainties due to the details of the fit procedure
are estimated by repeating
the fit with appropriate modifications. Different fit ranges (4
ranges) and different orders of the polynomial (1$^{\rm st}$ and
2$^{\rm nd}$ orders) are used in the fits. For $J/\psi \to
\gamma\gamma$, the uncertainty is estimated by averaging the
differences of the obtained yields with respect to the values derived
from the standard fit. For $J/\psi \to \gamma\phi$, the uncertainty is
estimated as the maximum difference between the obtained upper limits
and the upper limit derived from the standard fit. The uncertainties
from fitting are estimated as 2.7\% and 1.5\%, respectively.

The branching fractions for $\psi(3686) \to J/\psi \pi^+\pi^-$ and $\phi \to K^+K^-$
decays are taken from the PDG~\cite{pdg14}. The uncertainties of the branching fractions
are taken as systematic uncertainties in the measurements, which are 1.2\% and 1.0\%,
respectively.

The uncertainty in the number of $\psi(3686)$ events is 0.81\%, which
is measured by inclusive hadronic decays~\cite{npsip}.

Adding the uncertainties in quadrature yields total systematic uncertainties of
6.3\% and 10.0\% for $J/\psi \to \gamma\gamma$ and $J/\psi \to \gamma\phi$, respectively.

\begin{table}[htbp]
\begin{center}
\caption{Summary of the systematic uncertainties (\%). }\label{tab:syserr}
\begin{small}
\begin{tabular}{lcc} \hline
Sources            & $J/\psi \to \gamma\gamma$  & $J/\psi \to \gamma\phi$    \\ \hline
Tracking           & 4.0                          & 8.0  \\
Kaon identification  & -                          & 4.0  \\
Photon detection    & 0.5                  & 0.3 \\
Kinematic fit       & 1.9                        & 3.5  \\
Number of photons   & 3.0                   & - \\
$\cos\theta_{\pi^+\pi^-}$ requirement & 0.9             & 0.8 \\
$M^{\rm rec}_{\pi^+\pi^-}$ requirement  & -     & 1.4 \\
Fitting            & 2.7                        &  1.5 \\
$\mathcal{B}(\psi(3686) \to J/\psi \pi^+\pi^-) $ & 1.2 & 1.2 \\
$\mathcal{B}(\phi\to K^+K^-)$  & -              & 1.0 \\
Number of $\psi(3686)$   & 0.8                       & 0.8  \\ \hline
Total              & 6.3                        & 10.0 \\ \hline
\end{tabular}
\end{small}
\end{center}
\end{table}

\section{ RESULTS}
\label{sec:results}

Since no significant signals are observed, the upper limits on the branching fractions are determined by
\begin{equation}\label{eq:bfup}
\mathcal{B}(J/\psi \to f)^{\rm } <
\frac{ N^{\rm up}_{\rm sig} }{ N^{\rm tot}_{\psi(3686)} \times \epsilon \times \mathcal{B}_i \times (1-\Delta_{\rm sys}) },
\end{equation}
where $N^{\rm up}_{\rm sig}$ is the upper limit on the number of observed events for the
signal channel; $f$ represents $\gamma\gamma$ or $\gamma\phi$; $\epsilon$ is the detection efficiency determined
by MC simulation; $N^{\rm tot}_{\psi(3686)}$ is the total number of $\psi(3686)$ events,
$(106.41\pm0.86)\times 10^6$;
$\mathcal{B}_i$ denotes the branching fractions involved (such as $\mathcal{B}(\psi(3686) \to J/\psi\pi^+\pi^-) =(34.0\pm0.4)\%$
and $\mathcal{B}(\phi \to K^+K^-) = (48.9\pm0.5)\%$)~\cite{pdg14}; $\Delta_{\rm sys}$ is the
total systematic uncertainty, and $1/(1-\Delta_{\rm sys})$ is introduced to
estimate a conservative upper limit on the branching fraction. The individual values are summarized in Table~\ref{tab:results}.

Inserting $N^{\rm up}_{\rm sig}$, $ N^{\rm tot}_{\psi(3686)}$, $\epsilon$, $\mathcal{B}_i $
and $\Delta_{\rm sys}$ into Eq.(\ref{eq:bfup}), we obtain
\begin{equation}
\mathcal{B}(J/\psi \to \gamma\gamma)  < 2.7 \times 10^{-7} \nonumber
\end{equation}
and
\begin{equation}
\mathcal{B}(J/\psi \to \gamma\phi)  < 1.4 \times 10^{-6}. \nonumber
\end{equation}

\begin{table}[htbp]
\begin{center}
\caption{\small Results for both channels. }\label{tab:results}
\begin{small}
\begin{tabular}{llllll}\hline
                   & $\gamma\gamma$  & $ \gamma\phi$    \\ \hline
$N^{\rm obs}$      & $29.2\pm7.1$  & $0.0\pm4.6$  \\
$N^{\rm bkg}$      & $46.5\pm2.5$  & negligible  \\
$N^{\rm up}_{\rm sig}$(90\% C.L.)       & 2.8  & 6.9  \\
$\epsilon$ (\%)    & $30.72\pm0.07$  & $30.89\pm0.07$  \\
$ \mathcal{B}(J/\psi \to) $   (this work) & $<2.7\times 10^{-7}$ & $<1.4\times 10^{-6}$ \\
$ \mathcal{B}(J/\psi \to) $  (PDG~\cite{pdg14})    & $<50\times 10^{-7}$ &  - \\
\hline
\end{tabular}
\end{small}
\end{center}
\end{table}

\section{SUMMARY}

In this paper, we report on searches for $J/\psi \to \gamma\gamma$ and $J/\psi \to \gamma\phi$.
No significant signal is observed. We set the upper limits $\mathcal{B}(J/\psi \to \gamma\gamma)
< 2.7 \times 10^{-7}$ and $\mathcal{B}(J/\psi \to \gamma\phi)  < 1.4 \times 10^{-6}$ at the 90\% C.L. for
the branching fractions of $J/\psi$ decays into $\gamma\gamma$ and $\gamma\phi$, respectively.
The upper limit on
$\mathcal{B}(J/\psi \to \gamma\gamma)$ is one order of magnitude more
stringent than the previous upper limit, and $\mathcal{B}(J/\psi \to \gamma\phi)$ is the first
upper limit for this channel. Our results are consistent with C-parity conservation of the EM interaction.

\section{Acknowledgements}

The BESIII collaboration thanks the staff of BEPCII and the IHEP computing center for their strong support. This work is supported in part by National Key Basic Research Program of China under Contract No. 2015CB856700; National Natural Science Foundation of China (NSFC) under Contracts Nos.~10935007, 11121092, 11125525, 11235011, 11322544, 11335008; Joint Funds of the National Natural Science Foundation of China under Contracts Nos.~11079008, 11179007, U1232201, U1332201; the Chinese Academy of Sciences (CAS) Large-Scale Scientific Facility Program; CAS under Contracts Nos.~KJCX2-YW-N29, KJCX2-YW-N45; 100 Talents Program of CAS; German Research Foundation DFG under Contract No. Collaborative Research Center CRC-1044; Istituto Nazionale di Fisica Nucleare, Italy; Ministry of Development of Turkey under Contract No.~DPT2006K-120470; Russian Foundation for Basic Research under Contract No.~14-07-91152; U. S. Department of Energy under Contracts Nos.~DE-FG02-04ER41291, DE-FG02-05ER41374, DE-FG02-94ER40823, DESC0010118; U.S. National Science Foundation; University of Groningen (RuG) and the Helmholtzzentrum fuer Schwerionenforschung GmbH (GSI), Darmstadt; WCU Program of National Research Foundation of Korea under Contract No. R32-2008-000-10155-0


\begin{thebibliography}{99}

\bibitem{pdg14} J.~Beringer {\it et al.} [Particle Data Group], Phys.\ Rev.\ D {\bf 86}, 010001 (2012),
and 2013 partial update for the 2014 edition.
\bibitem{jpsic1} W.~Bartel {\it et al.}, Phys.\ Lett.\ B {\bf 66}, 489 (1977).
\bibitem{jpsic2} M.~Ablikim {\it et al.} [BES Collaboration], Phys.\ Rev.\ D {\bf 76}, 117101 (2007).
\bibitem{jpsic3} K.~Abe {\it et al.}, [Belle Collaboration], Phys.\ Lett.\ B {\bf 662}, 323 (2008).
\bibitem{jpsic4} G.~S.~Adams {\it et al.} [CLEO Collaboration], Phys.\ Rev.\ Lett.\ {\bf 101}, 101801 (2008).
\bibitem{npsip} M.~Ablikim {\it et al.} [BESIII Collaboration], Chin.\ Phys.\ C {\bf 37}, 063001 (2013).
\bibitem{continuumdata} M.~Ablikim {\it et al.} [BESIII Collaboration], Chin.\ Phys.\ C {\bf 37}, 123001 (2013).
\bibitem{bes3} M.~Ablikim {\it et al.} [BES Collaboration], Nucl.\ Instrum.\
Meth.\ Phys.\ Res.\ A {\bf 614}, 345 (2010).
\bibitem{bepc2} J.~Z.~Bai {\it et al.} [BES Collaboration], Nucl.\ Instrum.\
Meth.\ Phys.\ Res.\ A {\bf 344}, 319 (1994); {\bf 458}, 627 (2001).
\bibitem{bes3physics} Special issue on Physics at BES-III, edited by K. T. Chao and Y. F. Wang, Int. J.\ Mod.\ Phys.\ A {\bf 24} Supp. (2009).
\bibitem{boost} Z.~Y.~Deng {\it et al.}, High Energy Physics \& Nuclear Physics {\bf 30}, 371 (2006).
\bibitem{kkmc} S.~Jadach, B.~F.~L.~Ward, and Z.~Was, Comput. Phys. Commun. {\bf 130}, 260 (2000); Phys.\ Rev.\ D {\bf 63}, 113009 (2001).
\bibitem{evtgen} R.~G.~Ping, Chin.\ Phys.\ C {\bf 32}, 599 (2008); D.~J.~Lange, Nucl.~Instr.~Meth.~A {\bf 462}, 152 (2001).
\bibitem{pdg10} K.~Nakamura {\it et al.} [Particle Data Group], J.\ Phys.\ G {\bf 37}, 075021 (2010).
\bibitem{lundcharm} J.~C.~Chen, G.~S.~Huang, X.~R.~Qi, D.~H.~Zhang and Y.~S.~Zhu, Phys.\ Rev.\ D {\bf 62}, 034003 (2000).
\bibitem{bes2jpipi} J.~Z.~Bai {\it et al.} [BES Collaboration], Phys.\ Rev.\ D {\bf 62}, 032002 (2000).
\bibitem{bayesian} Y.~S.~Zhu, Chin.\ Phys.\ C {\bf 32}, 363 (2008).
\bibitem{feldman} G.~J.~Feldman and R. D. Cousins, Phys.\ Rev.\ D {\bf 57}, 3873 (1998).
\bibitem{trkpid} M.~Ablikim {\it et al.} [BESIII Collaboration], Phys.\ Rev.\ D {\bf 83}, 112005 (2011).
\bibitem{photon} M.~Ablikim {\it et al.} [BESIII Collaboration], Phys.\ Rev.\ Lett.\ {\bf 105}, 261801 (2010).


\end{thebibliography}
\end{document}